\documentclass{aa}

\usepackage{natbib,amsmath,amssymb,aasmacros,graphicx,url}

\newcommand\gr{$\gamma$-ray}

\newcommand\fermi{\textit{Fermi}}
\newcommand\fermilat{\textit{Fermi}-LAT}

\begin{document}

\title{Discovery of transient high-energy \gr\ emission from the BL Lacertae object 5C 3.178}

\titlerunning{HE \gr\ emission from 5C 3.178}
\authorrunning{Hugh Dickinson\and Christian Farnier}

\author{Hugh Dickinson\inst{\ref{inst1},\ref{inst2}}\and Christian Farnier\inst{\ref{inst1},\ref{inst2}}}
\institute{Oskar Klein Centre for Cosmoparticle Physics, AlbaNova, SE-106 91 Stockholm, Sweden\label{inst1}
\and
Department of Physics, Stockholm University, AlbaNova, SE-106 91 Stockholm, Sweden\label{inst2}
}

\abstract{We report the serendipitous discovery of transient, point-like high energy \gr\ emission coincident with the position of the suspected BL Lac object \object{5C 3.178}. The source was detected using the \fermi\ Large Area Telescope (LAT) at a significance level of $\sim8\sigma$ during a 200 day period which began on November 17th 2011 (MJD 55882). The observed \gr\ emission is characterised by a moderate 0.2-300 GeV flux $F_{0.2-300\,{\rm GeV}}=(8.22\pm2.04)\times10^{-9}$ ph cm$^{-2}$s$^{-1}$ and a hard power law spectrum with spectral index $\Gamma=1.76\pm0.09$. These properties appear consistent with the known sub-population of TeV \gr-emitting BL Lac objects, implying that the source may be detectable using atmospheric Cherenkov telescope arrays. Moreover, the temporal variability of the source suggests that real-time searches of the \fermilat\ all-sky dataset for weak emission on $\sim200$ day timescales may represent a rewarding strategy.}

\maketitle

\section{Introduction}\label{sec:intro}

Active Galactic Nuclei (AGNs) are compact, energetic astrophysical phenomena associated with the cores of many galaxies, which are paradigmatically assumed to be powered by disk accretion onto a supermassive ($M\gtrsim10^{8}M_{\odot}$) black hole. A sizeable fraction of AGNs are also launch sites for relativistic jets of plasma, which emit broad-band radiation with non-thermal energy spectra that extend from radio to \gr\ frequencies \citep[e.g.][]{1995PASP..107..803U}. 

BL Lacertae objects (hereafter BL Lacs) are a subset of AGNs that have jet axes that are oriented close to the observer's line of sight. The observed spectral energy distributions (SEDs) of BL Lacs are dominated by relativistically beamed emission from charged particles entrained in the jets and exhibit a characteristic two-component structure. The individual components are visible as low and high energy peaks in a $\nu F_{\nu}$ representation of the SED and are typically identified with distinct radiative processes occurring within the jets. The low energy component is usually attributed to synchrotron emission while inverse-Compton up-scattering of this synchrotron emission or other ambient soft radiation fields is assumed to dominate at higher energies.

BL Lacs are an established class of variable high energy (HE; $10^{-3}\lesssim E_{\gamma}/{\rm GeV}\lesssim50$) and very high energy (VHE;  $10^{-3}\lesssim E_{\gamma}/{\rm TeV}\lesssim10^{2}$) \gr\ sources. Indeed, the second \fermilat\ (2FGL) source catalogue includes 435 confirmed BL Lac objects, of which 36 have also been detected at TeV energies \citep{2012ApJS..199...31N}.
In a recent review, \citet{2013ApJ...764..119S} examined the \fermilat\ light curves of 12 established VHE \gr-emitting BL Lacs, which reveal significant flux variability on timescales ranging from days to months. Furthermore, a population study of 395 BL Lac objects listed in the second \fermilat\ AGN catalogue (2LAC) also revealed clear evidence for HE variability, although it is apparently less prevalent in sources with hard \gr\ spectra \citep{2011ApJ...743..171A}.

\object{5C 3.178} is a point-like radio source located to the southeast of M31 at $(\alpha_{\rm J2000}, \delta_{\rm J2000} )  = (00^{\rm h}47^{\rm m}55^{\rm s}.220, = +39^{\circ}48'57''.60)$ \citep{2011NewA...16..503M}. The source was initially discovered using the Cambridge one-mile radio telescope \citep{1969MNRAS.144..101P} and has been provisionally identified as a BL Lac object based on the absence of detectable emission features in its optical spectrum \citep{1987AJ.....93....1B,1995ApJS..101..255D}. Multi-wavelength counterparts have been identified at near infra-red \citep{2006AJ....131.1163S, 2011NewA...16..503M}, optical \citep{1995ApJS..101..255D}, ultraviolet \citep{2009MNRAS.399..728F} and X-ray \citep{1999A&A...349..389V,2008A&A...480..611S} frequencies. A source redshift of $z=0.2517\pm0.001$ has been established on the basis of optical absorption features imprinted by the host galaxy \citep{1995ApJS..101..255D}.

The Large Area Telescope is the principal scientific instrument on the \fermi\ Gamma Ray Space Telescope spacecraft. It is a pair conversion telescope with a $\sim1$ m effective area, which is sensitive to photons with energies between $20$ MeV and $300$ GeV. The field of view of the \fermilat\ encompasses 2.4 steradians, and the instrument typically operates in a continuous survey mode, providing all-sky coverage every two orbits. The operation and performance of the \fermilat\ are described in detail by \citep{2009ApJ...697.1071A}.

This article describes the discovery and analysis of a point-like \gr\ signal coincident with the position of \object{5C 3.178}, which was serendipitously identified while searching unsuccessfully for HE emission associated with a 200 day flare of the nearby ultra-luminous x-ray source (ULX) \object{XMMU J004243.6+412519} \citep{2013Natur.493..187M}. Section \ref{sec:analysis} describes details of the data selection and analysis procedure which were employed to produce the results presented in section \ref{sec:results}. Section \ref{sec:disc} discusses the implications of this detection in the context of HE and VHE \gr\ emission from BL Lac objects.

\section{Data Analysis}\label{sec:analysis}

The \fermilat\ data were analysed using version \texttt{v9r27p1} of the Fermi Science Tools\footnote{\url{http://fermi.gsfc.nasa.gov/ssc/data/analysis/scitools/overview.html}} in conjunction with the \texttt{P7SOURCE\_V6} (Pass 7) instrument response functions. The dataset under consideration was extracted from a circular region of interest (ROI) of radius 10 degrees centred on \object{5C 3.178} and comprised all \texttt{SOURCE} class photon-like events detected by the \textit{Fermi}-LAT before MJD 56282 (2012-12-21) with energies between 200 MeV and 300 GeV. Good time intervals (GTIs) were generated using the recommended selection expression\footnote{\texttt{(DATA\_QUAL==1) \&\& (LAT\_CONFIG==1) \&\& ABS(ROCK\_ANGLE)<52}} and ROI-based maximum zenith angle cut\footnote{This cut excludes all time intervals when the zenith angle of any part of the ROI exceeds $100^{\circ}$.}, before being applied to isolate subsections of the data for subsequent analysis. 

The reduced data were analysed using an un-binned maximum likelihood approach \citep{1996ApJ...461..396M} implemented by the \texttt{gtlike} utility\footnote{\url{http://fermi.gsfc.nasa.gov/ssc/data/analysis/scitools/help/gtlike.txt}}, in which two alternative model hypotheses were compared by maximising their respective likelihoods with respect to the the observed photon distribution. As a nominal \textit{null} hypothesis, the expected \gr\ signal within the ROI was modelled using a combination of the standard galactic (\texttt{gal\_2yearp7v6\_v0}) and isotropic (\texttt{iso\_p7v6source}) diffuse emission models. The model also incorporated all point-like \gr\ emitters within $20^{\circ}$ of \object{5C 3.178} that are listed in the 2FGL catalogue. To prevent genuine variability or statistical fluctuations of the signals from  nearby \gr\ emitters from affecting the analysis, the overall flux normalisations of sources within a $10^{\circ}$ radius of the target were treated as variable parameters. In order to ameliorate any local mis-modelling of the diffuse \gr\ emission within the ROI, the overall normalisations of the isotropic and galactic template components were also allowed to vary during the likelihood analysis. To limit the number of degrees of freedom in the final likelihood, the remaining source model parameters were fixed to the best fitting values published in the 2FGL. A second, \textit{alternative} hypothesis included a model for putative emission from an additional point-like \gr\ source with a power-law spectral shape parameterised as
\begin{equation}
dN/dE=N_{0}(E/E_{0})^{-\Gamma}.
\end{equation}
The normalisation ($N_{0}$) and spectral index ($\Gamma$) were allowed to vary during the likelihood optimisation, while the energy scale ($E_{0}$) was fixed to 1 GeV.  

Evidence for the detection of an additional HE \gr\ source was evaluated in terms of a likelihood ratio test statistic 
\begin{equation} 
TS=-2\ln\frac{L_{\max, 0}}{L_{\max,1}}
\end{equation}
where $L_{\max, 0}$ and $L_{\max, 1}$ are the maximum likelihood values obtained when fitting the observed data using the null and alternative hypothesis models respectively. If the null hypothesis is true, then $\sqrt{TS}$ is approximately equivalent to the source detection significance in the framework of standard normal theory. 

\section{Results}\label{sec:results}

\begin{figure*}[htbp]
\begin{center}
\includegraphics[width=0.5\textwidth]{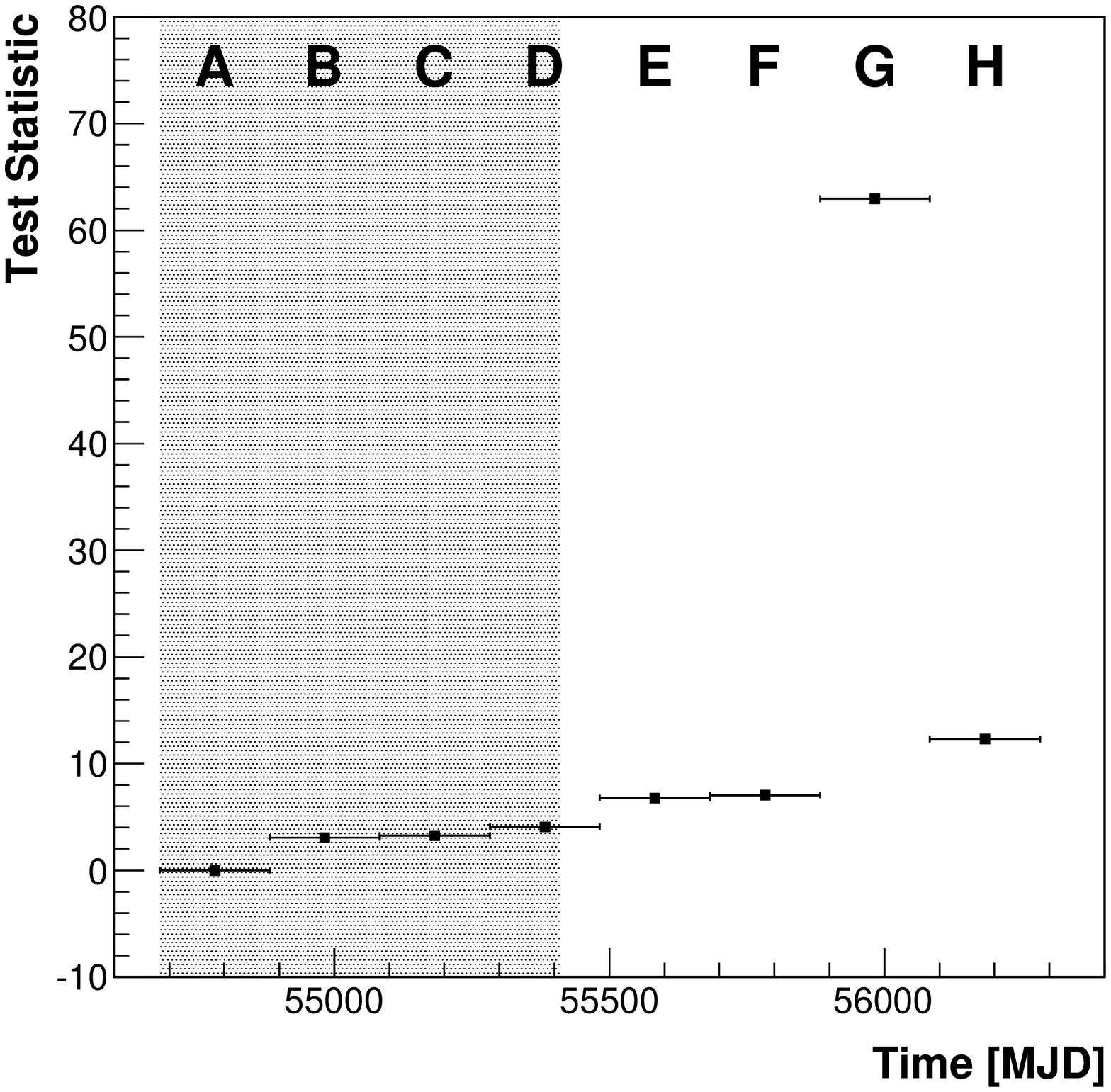}\includegraphics[width=0.5\textwidth]{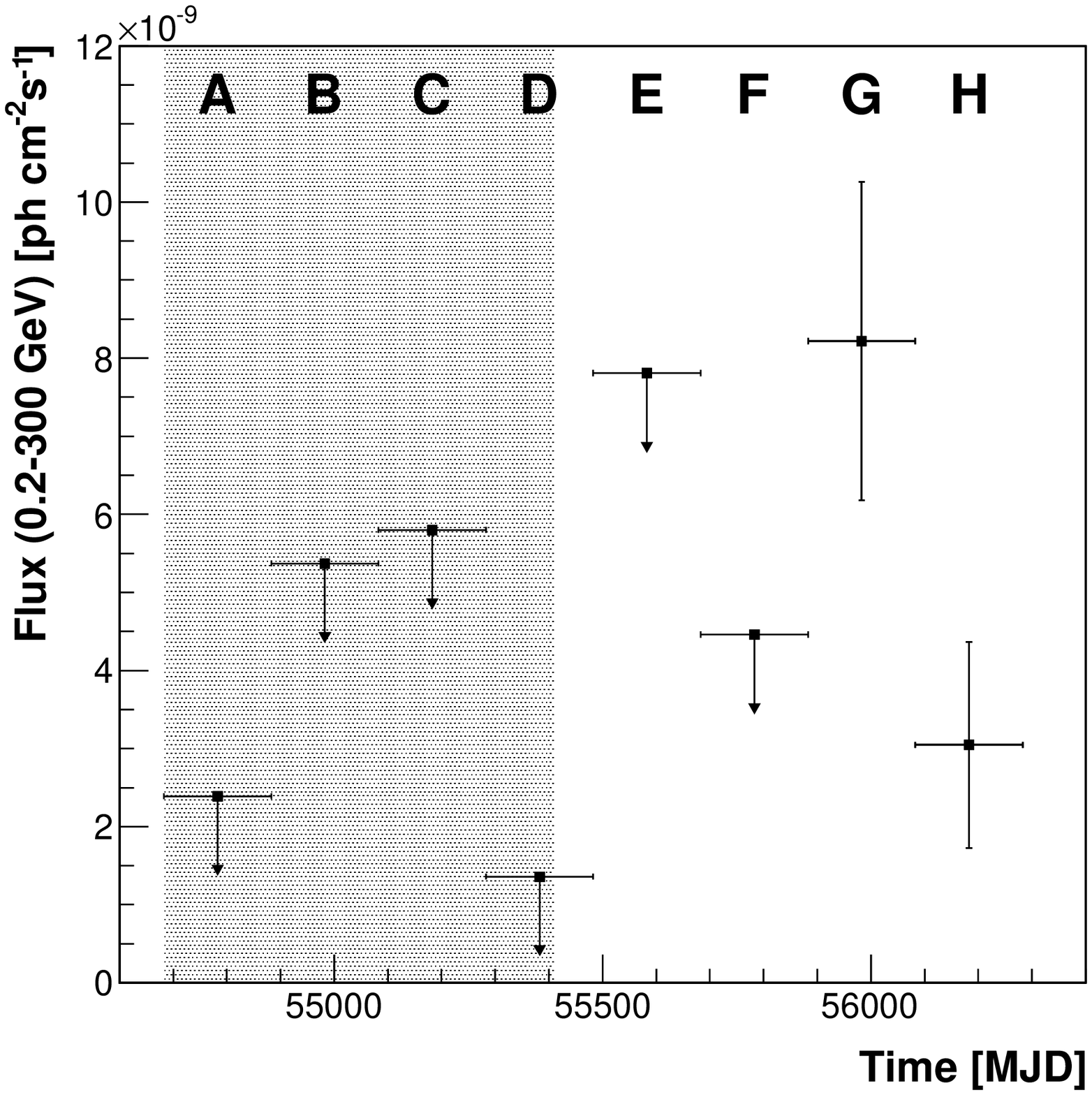}
\caption{\textit{Left panel:} Values of $TS$ corresponding to intervals A-H. \textit{Right panel:}. 95\% confidence upper limits on the 0.2-300 GeV \gr\ flux are presented for intervals during the intervals in which the source was not detected. The shaded areas in both plots indicate the temporal extent of the \fermilat\ dataset used to compile the 2FGL catalogue \citep{2012ApJS..199...31N}.}
\label{fig:lightcurve}
\end{center}
\end{figure*}

As explained in $\S$\ref{sec:intro}, the initial source detection corresponded to \fermi\ data that were obtained during the 200 day flaring cycle of the ULX XMMU J004243.6+412519. A subsequent analysis of the complete \fermilat\ dataset revealed strong evidence ($TS=54.1\approxeq 7.4\sigma$) for a spectrally hard ($\Gamma\approx1.7$) point-like HE \gr\ source close to the position of \object{5C 3.178}, with a time-averaged 0.2-300 GeV flux of $(1.78\pm0.79)\times10^{-9}$ ph cm$^{-2}$s$^{-1}$. Given its apparent brightness, the absence of this source from both the first and second \fermilat\ source catalogues \citep{2010ApJS..188..405A,2012ApJS..199...31N} suggests a transient origin for the observed \gr\ emission. In the absence of any \textit{a-priori} expectation for likely variability timescales, the overall dataset was subdivided into 200 day intervals (labelled A-H) to match the duration of the initial detection dataset. Each interval was individually re-analysed to obtain the results listed in Table \ref{tab:results} and illustrated in Figure \ref{fig:lightcurve}.

Assuming that the test statistic is approximately $\chi^{2}_{1}$ distributed \citep{1996ApJ...461..396M}, then the probability of obtaining a value of $TS$ which exceeds some threshold $TS_{0}$ in a \textit{single trial} is
\begin{equation}
P_{1}(TS \geq TS_{0}) = \int_{TS}^{\infty}{\chi^{2}_{1}(x)dx}\\
\end{equation}
Under the reasonable assumption that the measured \gr\ signals corresponding to non-overlapping time slices are statistically independent, then the probability of obtaining a $TS>TS_{0}$ \textit{at least once} after testing all 8 sub-intervals is 
\begin{equation}
P_{8} = \displaystyle\sum_{i=1}^{8}{\binom{8}{1}P_{1}^{i}(1-P_{1})^{8-i}}\\
\end{equation}
Defining $\Phi^{-1}$ to be the complement of the quantile of the $\chi^{2}_{1}$ distribution, $P_{8}$ can be used to derive a trials-corrected test statistic   
\begin{equation}
TS'=\Phi^{-1}(P_{8}).
\end{equation}

A trials-corrected test statistic value of $TS'=58.9$ ($\approxeq 7.7\sigma$) provides clear evidence for transient \gr\ emission during interval G, which begins on MJD 55882 (2011-09-17). Further subdivision of interval G into 10 week, 2 week and 1 week time slices does not reveal significant flux variability on shorter timescales. The source was not conclusively detected during any other interval, although there is marginal evidence for emission ($TS'\approx8.5$) during interval H. For intervals yielding $TS<9$, Table \ref{tab:results} lists 95\% confidence level upper limits on the  0.2-300 GeV flux, which are also plotted in Figure \ref{fig:lightcurve}. The time-resolved analysis also reveals that the source would not have been detectable during the first 24 months of the \fermi\ mission, which explains its absence from the \fermilat\ source catalogues.
\begin{table*}
\begin{center}
\begin{tabular}{llllllll}
\hline\hline
Interval & $T_{\rm start}$ & $T_{\rm end}$ & $T_{\rm start}$ & $TS$  		& $TS'$ 		& Flux (0.2-300 GeV) 	& $\Gamma$ \\ 
&(MET)		& (MET) &(UTC)		&	& 	& ($10^{-9}$ ph cm$^{-2}$s$^{-1}$) 	&\\
\hline
A& 239557417	&	256837417  & 2008-08-04 15:43:36 &	0.0	&0.0&$<2.39$&\ldots\\
B& 256837417	&	274117417  & 2009-02-20 15:43:35 & 3.0	& 0.4 &$<5.37$&\ldots\\
C& 274117417	&	291397417  & 2009-09-08 15:43:35	& 3.3	& 0.6&$<5.80$&\ldots\\
D& 291397417	&	308677417  & 2010-03-27 15:43:35 &4.1	& 1.0&$<1.36$&\ldots\\
E& 308677417	&	325957417  & 2010-10-13 15:43:35 & 6.8	& 3.2&$<7.81$&\ldots\\
F& 325957417	&	343237417  & 2011-05-01 15:43:35	&7.1	& 3.5&$<4.46$&\ldots\\
G& 343237417	&	360517417  & 2011-11-17 15:43:35	&63.0	& 58.9 &$8.22\pm2.04$&$1.76\pm0.09$\\
H& 360517417	&	377797417  & 2012-06-04 15:43:35	&12.3	& 8.5 &$3.05\pm1.32$&$1.69\pm0.12$\\
All& 239557417 & 377797417	& 2008-08-04 15:43:36 &54.1	& \ldots &$1.78\pm0.79$&$1.70\pm0.15$\\
\hline
\end{tabular}
\end{center}
\caption{Analysis results corresponding to eight 200 day subintervals (A-H) as well as the complete \fermilat\ dataset. Test statistic values are quoted before ($TS$) and after ($TS'$) correcting for the number of independent trials. For intervals during which the source was detected such that $TS>9$, the HE \gr\ flux and best fitting power-law spectral index ($\Gamma$) are also listed. For other intervals, upper limits to the HE \gr\ flux (at the 95\% confidence level) are presented.}\label{tab:results}
\end{table*}

Given the limited duration of the detectable \gr\ emission, subsequent analyses only considered the restricted dataset of events detected during interval G, in order to maximise the signal-to-noise ratio. The \texttt{gtfindsrc} utility\footnote{\url{http://fermi.gsfc.nasa.gov/ssc/data/analysis/scitools/help/gtfindsrc.txt}} was applied to derive the most likely coordinates of the observed \gr\ emission, yielding $(\alpha_{\rm LAT, J2000}, \delta_{\rm LAT, J2000}) =  (00^{h}48^{m}20^{s}.89, +39^{\circ}52'26''.1)$ with a 95\% confidence error circle radius of $0.06^{\circ}$. Using these coordinates, the automatic source association method described by \citep{2010ApJS..188..405A} and implemented by the ScienceTools executable \texttt{gtsrcid}, was applied to search for plausible multi-wavelength counterparts to the observed HE \gr\ emitter. After considering all sources listed within the catalogues specified in Table 9 of \citet{2012ApJS..199...31N}, the procedure identified \object{5C 3.178} as the only viable candidate, deriving a posterior probability of 91\% that the association is correct.

\begin{figure*}[htbp]
\begin{center}
\includegraphics[width=0.5\textwidth]{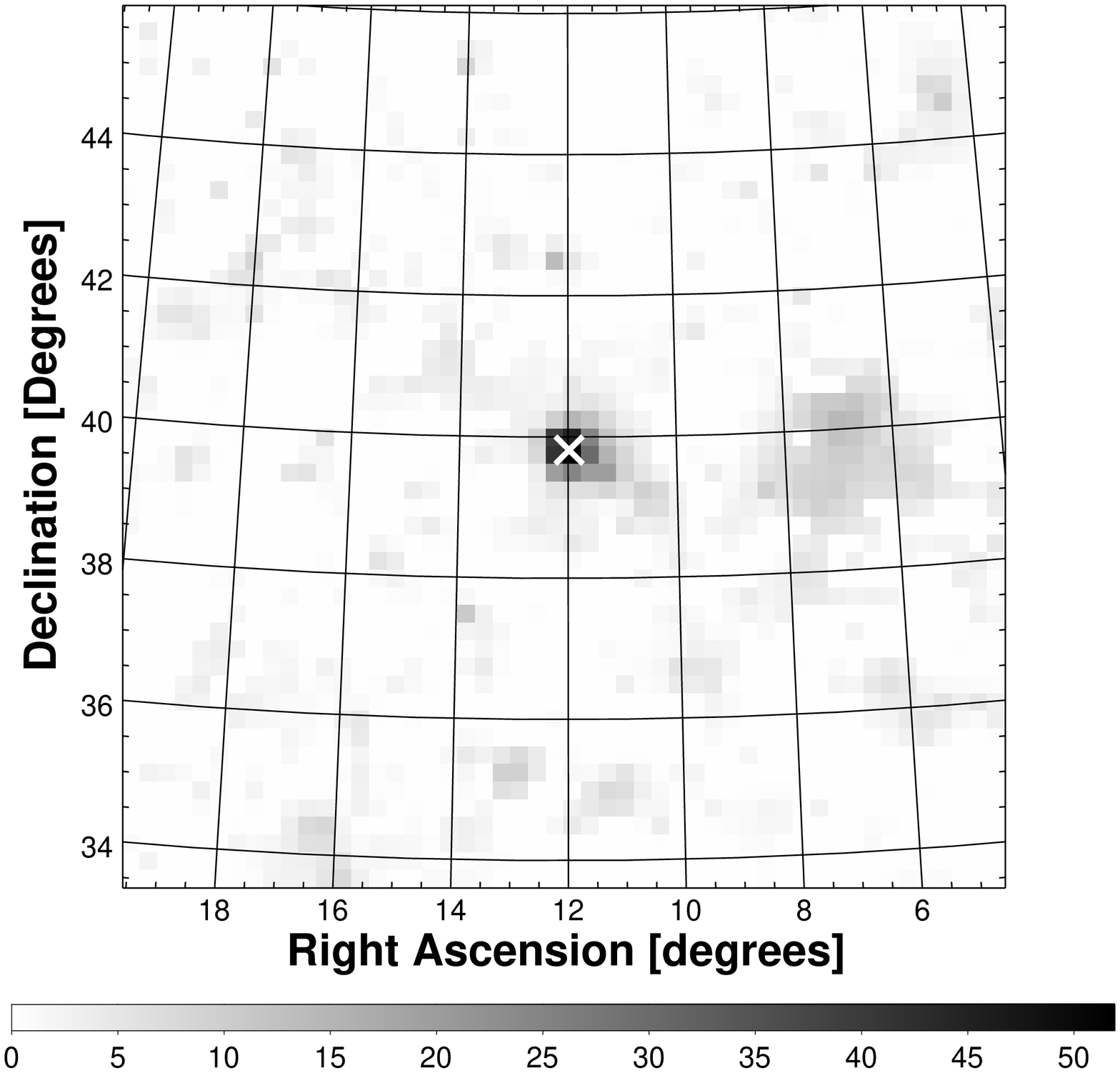}\includegraphics[width=0.5\textwidth]{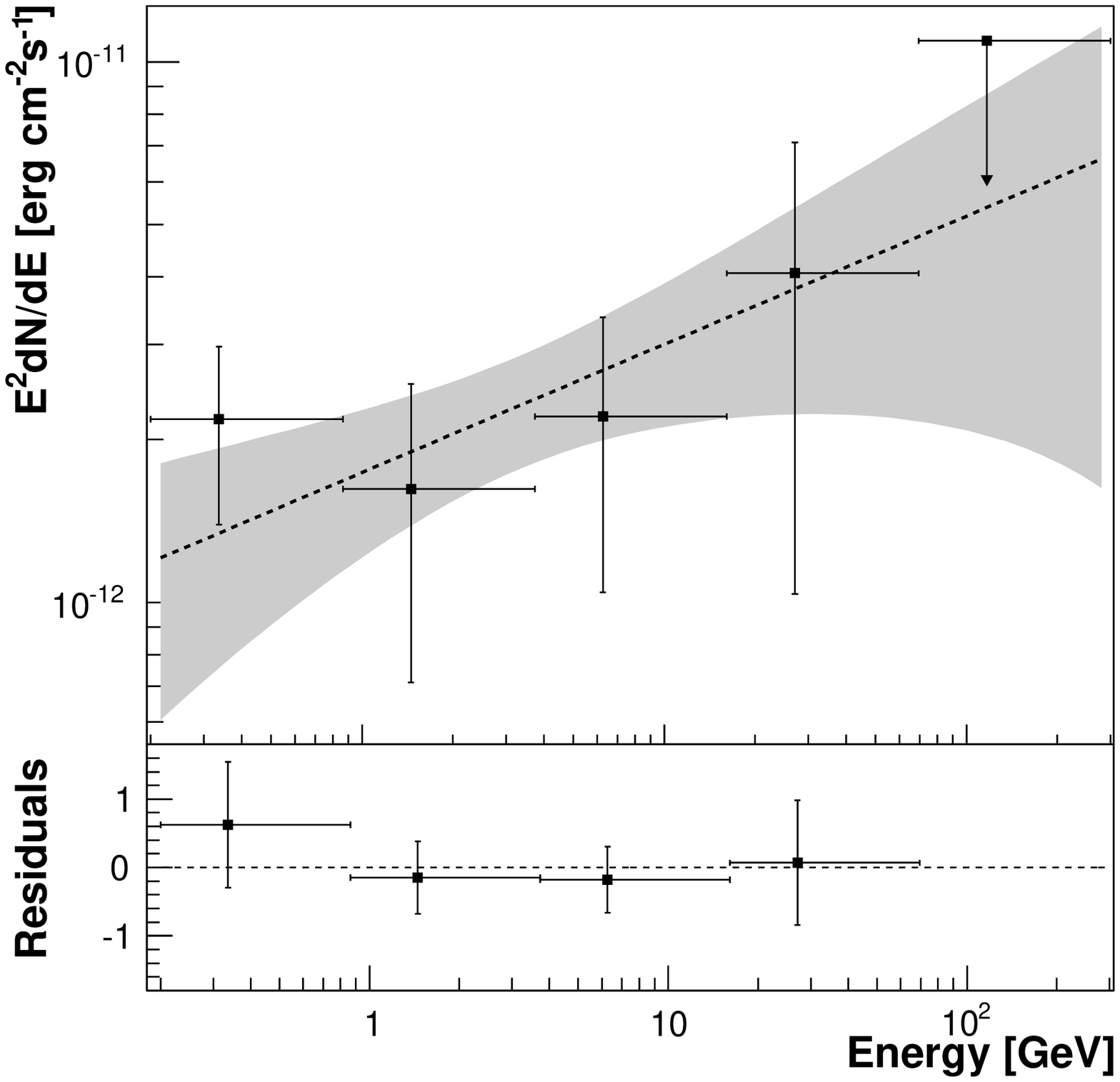}
\caption{\textit{Left panel:} Unbinned likelihood test statistic map corresponding to subinterval G of the \fermilat\ dataset. The white cross marks the nominal radio position of \object{5C 3.178}. \textit{Right panel:} HE \gr\ spectral energy distribution between 0.2 and 300 GeV corresponding to subinterval G of the \fermilat\ dataset. The dotted line indicates the best-fitting power-law model, and the grey region indicates the $1\sigma$ fit uncertainties. Residuals of the observed data points with respect to the best fitting power law spectral model are also presented.}
\label{fig:sed_and_ts}
\end{center}
\end{figure*}

The \fermilat\ data corresponding to subinterval G were used in conjunction with the \texttt{gttsmap} tool\footnote{\url{http://fermi.gsfc.nasa.gov/ssc/data/analysis/scitools/help/gttsmap.txt}} to generate the $TS$ sky map presented in the left panel of Figure \ref{fig:sed_and_ts}. To generate this map, the null hypothesis model described in $\S$\ref{sec:analysis} was used, with all parameters except for the diffuse component normalisations fixed to the best fitting values that were obtained for this subinterval during the likelihood analysis. The transient \gr\ source is clearly visible as a point-like enhancement of the computed $TS$ values, coincident with the nominal position of \object{5C 3.178} \citep{2011NewA...16..503M}. 

The observed 0.2-300 GeV spectral energy distribution and best-fitting power law model corresponding to interval G are presented in the right-hand panel of Figure \ref{fig:sed_and_ts}. The data correspond well with the assumed model and indicate that the transient \gr\ emission is spectrally hard, with a power law index $\Gamma_{G}=1.76\pm0.09$.

\section{Discussion}\label{sec:disc}

The results presented in $\S$\ref{sec:results} provide clear evidence for transient, spectrally hard HE emission from a point-like \gr\ source that is positionally coincident with the BL Lac object \object{5C 3.178}.
The spectral properties and temporal variability of the transient are both consistent with the characteristics of other BL Lac objects that have previously been detected by the \fermilat\ \citep{2011ApJ...743..171A,2013ApJ...764..119S}. On the basis of this combined evidence, it seems reasonable to attribute the observed \gr\ emission to a HE flare of the BL Lac object \object{5C 3.178}.

The hard spectral index of the observed HE emission is also consistent with the sub-population of 36 High-Synchrotron-Peaked BL Lac objects (HBLs) that have also been detected at VHE \gr\ energies, all of which have $\Gamma<2.3$ \citep{2013ApJ...764..119S}. Indeed, the moderate redshift of \object{5C 3.178} suggests that the source may be a viable target for observations by  ground-based Cherenkov telescope arrays. Combined HE and VHE \gr\ observations HBLs have been used to probe the properties of the extragalactic background light \citep[e.g.][]{2012Sci...338.1190A,2013A&A...550A...4.} and can provide strong constraints on the value of bulk cosmological properties such as the strength of the extragalactic magnetic field \citep{2011A&A...529A.144T,2012ApJ...747L..14V}. 

Studies of this type invariably benefit from multi-wavelength observations and strictly simultaneous data are preferred when dealing with variable and unpredictable \gr\ signals. Without an \textit{a-priori} expectation of transient \gr\ emission, achieving this simultaneity requires rapid identification of potential targets in order to coordinate the instruments involved. Atmospheric Cherenkov telescope arrays such as MAGIC II \citep{2012APh....35..435A} or VERITAS \citep{2006APh....25..391H} provide good sensitivity above $\sim100$ GeV, but are characterised by limited fields-of-view ($\lesssim5^{\circ2}$) which renders the serendipitous discovery of flaring \gr\ sources using these instruments unlikely. In contrast, continuous survey observation mode of the \fermilat\ effectively guarantees that any transient emission lasting longer than $\sim6$ hours will at least be \textit{observed}, although limited sensitivity in the GeV energy range may lead to fainter sources escaping detection. 

Systematic blind searches for variable signals within the \fermilat\ all-sky dataset are extremely computationally intensive and can only consider a limited number preselected variability timescales. Currently, automatic monitoring for transient \gr\ signals in the latest \fermilat\ data is provided by the \textit{Fermi} Flare Advocate (FA) service \citep{2012AIPC.1505..697C}. This facility performs preliminary analyses that are optimised to search for bright ($F_{E>100\,{\rm MeV}}\gtrsim10^{-6}\,{\rm ph s}^{-1}{\rm cm}^{2}$) sporadic \gr\ emission within intervals lasting 6 hours, one day, and one week. Restricting transient searches in this manner inevitably sacrifices sensitivity to signals with longer variability timescales. Indeed, if the proposed association with \object{5C 3.178} is correct the \gr\ signal identified in this work may indicate a population of mildly variable BL Lac objects that are overlooked by the current FA policy. Assuming that this is the case, the temporal characteristics of the observed HE signal appear to motivate all-sky searches for moderately bright \gr\ emission that varies on $\sim100-200$ day timescales.

\section{Summary}\label{sec:summary}

\fermilat\ data have been used to identify a point-like source of transient HE \gr\ emission that is positionally coincident with the BL Lac object \object{5C 3.178}. The probability that \object{5C 3.178} is the correct association for the transient emission was calculated to be 91\%, while a search for alternative counterparts yielded no viable candidates. A temporally resolved analysis yielded a trials-corrected maximum likelihood ratio test statistic of 58.9 (equivalent to $\approx 7.7\sigma$) and localised the observed \gr\ signal to the time interval between MJD 55882 (2011-11-17) and MJD 56102 (2012-06-24). During this interval, the source was characterised by a hard power law energy spectrum ($dN/dE\propto E^{-\Gamma}$) with spectral index $\Gamma=1.76\pm0.09$ and a 0.2-300 GeV flux $F_{0.2-300\,{\rm GeV}}=(8.22\pm2.04)\times10^{-9}$ ph cm$^{-2}$s$^{-1}$. In combination, these spectral and temporal properties are reminiscent of a limited number of HBLs that have previously been observed by the \fermilat, and which are often VHE \gr\ emitters. The detection of \object{5C 3.178} using ground-based atmospheric Cherenkov telescope arrays would extend the existing catalogue of BL Lacs that have been detected at both GeV and TeV energies, enabling a more secure classification of the source and simultaneously establishing \object{5C 3.178} ($z\approx0.25$) as the fourth most distant object in this category\footnote{See e.g. \url{http://tevcat.uchicago.edu}}. As well as elucidating the intrinsic processes that operate within AGNs, simultaneous HE and VHE observations of distant HBLs can also be used to constrain the properties of extragalactic photon fields or the intergalactic magnetic field. 

\begin{acknowledgements}
The authors thank Jan Conrad and Stephan Zimmer for their valuable comments and advice during the preparation of this manuscript.
\end{acknowledgements}

\end{document}